\documentclass[%
preprint,
 amsmath,amssymb,
 aps,
]{revtex4-1}

\usepackage{graphicx}
\usepackage{dcolumn}
\usepackage{bm}
\usepackage{CJK}

\begin{document}
\begin{CJK*}{UTF8}{}


\title{No-Core Monte Carlo Shell Model Calculations with Unitary Correlation Operator Method and Similarity Renormalization Group}

\author{Lang Liu {\CJKfamily{gbsn}(刘朗)}}
\email{liulang@jiangnan.edu.cn}
\affiliation{School of Science, Jiangnan University.}%




\date{\today}

\begin{abstract}
The unitary correlation operator method (UCOM) and the similarity renormalization group theory (SRG) are compared and discussed in the framework of no-core Monte Carlo shell model (MCSM) calculations for $^{3}$H and $^{4}$He. The treatment of spurious center-of-mass motion by Lawson's prescription is performed in the MCSM calculations. These results with both transformed interactions show good suppression of spurious center-of-mass motion with proper Lawson's prescription parameter $\beta_{\rm c.m.}$ values. The UCOM potentials obtains faster convergence of total energy for the ground state than that of SRG potentials in the MCSM calculations, which differs from the cases in the no-core shell model calculations (NCSM). This differences are discussed and analyzed in terms of the truncation scheme in the MCSM and NCSM, as well as the properties of potentials of SRG and UCOM.
\end{abstract}

\maketitle
\end{CJK*}


\section{Introduction}

Nuclear \textit{ab initio} calculation is one of the most effective approaches for investigating the structure of nuclei. By using realistic nuclear interactions, \textit{ab initio} nuclear many-body calculations have been performed in the past decade. In Green's function Monte Carlo (GFMC) calculations the exact ground-state wave function is calculated by treating the many-body Green's functions in a Monte Carlo approach~\cite{Pieper2001,Pieper2002,Pieper2005}. The GFMC calculations of light nuclei up to $^{12}$C with the Argonne interaction reproduce the experimental nuclear binding energies and radii as well as the spectra. Another \textit{ab initio} approach for nuclei up to $A = 14$ is the no-core shell model (NCSM)~\cite{Navratil2000,Caurier2002,Navratil2009}. The Monte Carlo shell model (MCSM) has been introduced recently to study some light nuclei and might been considered as a new method to push limit of present \textit{ab initio} calculations because it reduces the dimension of basis dramatically compared with other shell model calculation~\cite{Liu2012,Shimizu2010}.

However, the straightforward application of the realistic interactions in nuclear many-body calculations is still difficult due to the strong short-range repulsion which generates strong correlations in the nuclear many-body state. The unitary correlation operator method (UCOM) is one of the methods used to tackle this problem by introducing a unitary transformation such that the transformed many-body states contain the information on the dominant correlations in the nuclear many-body system~\cite{Feldmeier1998,Neff2003,Roth2010}. The similarity renormalization group is another unitary transformation which aims at the pre-diagonalization of a matrix representation of the Hamiltonian in a chosen basis by means of a renormalization group flow evolution~\cite{Wegner1994,Wegner2000}.  It has been proved to be an effective approach not only for \textit{ab initio} calculations, but also in nuclear covariant density functional theory to investigate the symmetries of Dirac Hamiltonian very recently~\cite{Guo2014}.

It is interesting to make a comparison for the MCSM and NCSM with different transformed potentials: SRG-transformed and UCOM-transformed interactions in order to provide some benchmarks for the MCSM and NCSM. In Ref.~\cite{Roth2008}, the SRG-transformed and UCOM-transformed potentials in NCSM calculation are discussed from matrix elements to many-body calculations. In this work, we focus on the properties of SRG-transformed and UCOM-transformed potentials in the MCSM calculations for $^{3}$H and $^{4}$He. In Section~2, the theoretical framework for the MCSM is briefly outlined. The numerical details, results, and discussion of many-body calculation results are presented in Sec.~3. Finally, a brief summary is given in Sec.~4.

\section{Theoretical framework}
\label{sec:theory}

The main idea of the MCSM is to diagonalize the Hamiltonian in a subspace spanned by the MCSM basis, which is generated in a stochastic way.

We begin with the acting of imaginary-time evolution operator to a state $ |\Psi^{(0)}\rangle $
\begin{eqnarray}\label{eq3_2}
e^{-\beta H} |\Psi^{(0)}\rangle\,=\,\sum\limits_{i}e^{-\beta E_{i}}c_{i}|\psi_{i}\rangle,
\end{eqnarray}
where $H$ is a given general (time-independent) Hamiltonian and $\beta\propto T^{-1}$ is a real number with $T$ being analogous to a temperature. In Eq.~\ref{eq3_2}, $E_i$ is the $i$-th eigenvalue of $H$. $ |\psi_{i}\rangle $ is the corresponding eigenstate and $c_{i}$ its amplitude in the initial state. For $\beta$ large enough, only the ground and low-lying states survive. But the actual handling is very complicated for $H$ containing a two-body (or many-body) interaction.

The Hubbard-Stratonovich (HS) transformation~\cite{Hubbard1959,Stratonovich1957} can be used to ease the difficulty mentioned above. We then move to the formula
\begin{eqnarray}\label{eq3_17}
|\Phi(\sigma)\rangle\,\varpropto\,e^{-\beta h(\sigma)}|\Psi^{(0)}\rangle,
\end{eqnarray}
where $h(\sigma)$ is a one-body Hamiltonian obtained through the HS-transformation and $\sigma$ is a set of random numbers (auxiliary fields). The right-hand-side of this relation can be interpreted as a means to generate all basis vectors needed for describing the ground state and the low-lying states. For different values of the random variable, $ \sigma $, one obtains different state vectors, $ |\Phi(\sigma)\rangle $, by Eq.~(\ref{eq3_17}). These vectors are labeled as candidate states and selected as MCSM basis by a procedure of energy comparison.

During the MCSM generation of the basis vectors, symmetries, e.g. rotational and parity symmetry, are restored before the diagonalization as more basis vectors are included. All MCSM basis states are projected onto good parity and angular momentum quantum numbers by acting with the corresponding projection operators. We diagonalize the Hamiltonian in a subspace spanned by those projected basis vectors. The number of the MCSM basis states is referred to as the MCSM dimension. The basis generation process for general cases is outlined in Ref.~\cite{Otsuka2001}.

As more than one major shell is included in the MCSM calculation, the spurious center-of-mass motion must be taken into account. The Lawson'prescription is adopted to suppress the spurious center-of-mass motion in good approximation for major shell truncation~\cite{Gloeckner1974}. The total Hamiltonian can be separated into an intrinsic part and a center-of-mass part
\begin{eqnarray}\label{eq3_31}
H'\,=\,H_{int.}+\beta_{c.m.} H_{c.m.},
\end{eqnarray}
where $ H_{int.} $ is the intrinsic Hamiltonian. The $ H_{c.m.} $ is defined by
\begin{eqnarray}\label{eq3_32}
H_{c.m.}\,=\,\frac{\mathbf{P}^2}{2AM}+\frac{1}{2}MA\omega^2\mathbf{R}^2-\frac{3}{2}\hbar\omega,
\end{eqnarray}
where $\mathbf{R}$ and $\mathbf{P}$ are the coordinate and momentum of the center of mass, respectively. In general, by taking sufficiently large values of $ \beta_{c.m.} $, the spurious components become smaller and smaller for the low-lying eigenstates of $H'$. More details of the MCSM can be found in Refs~\cite{Liu2012,Otsuka2001}.

The basic idea of the SRG approach in the formulation of Wegner [11,12] is to transform the initial Hamiltonian $H$ of a many-body system into a diagonal form with respect to a given basis. The renormalization group flow equation governing the evolution of the Hamiltonian is of the form
\begin{eqnarray}
\dfrac{d\,H_{\alpha}}{d\,\alpha} = \left[ \eta_{\alpha},\ H_{\alpha}\right],
\end{eqnarray}
where $\alpha$ is the flow parameter and $H_{\alpha}$ the evolved Hamiltonian with $H_0 = H$. Analogous equations can be formulated for the operators of all observables one is interested in. In general terms the anti-hermitian generator $\eta_{\alpha}$ of the flow can be written as
\begin{eqnarray}
\eta_{\alpha} = \left[ \text{diag}(H_{\alpha}),\ H_{\alpha}\right],
\end{eqnarray}
where diag($H_{\alpha}$) refers to the diagonal part of the Hamiltonian in a given basis. This choice can be understood in intuitive terms: if the Hamiltonian commutes with its diagonal part with respect to a given basis, then the generator vanishes and the evolution has reached a fix point. More details of SRG can be found in Ref.~\cite{Wegner1994,Roth2008}.

The main idea of the UCOM can be interpreted by the expression
\begin{eqnarray}
\langle\Psi |H| \Psi' \rangle = \langle\Phi |C^{\dag}HC| \Phi' \rangle = \langle\Phi |\hat{H}| \Phi' \rangle,
\end{eqnarray}
where $|\Psi\rangle$ and $|\Psi'\rangle$ are correlated wave functions. $H$ is the nuclear Hamiltonian including realistic nucleon-nucleon interaction. $\hat{H}$ is a transformed potential by operator $C$. In the UCOM, $C = C_{r}\,C_{\Omega}$ indicates that $C$ is composed by central correlation operator $C_r$ and tensor correlation operator $C_{\Omega}$. More details can be found in Ref.~\cite{Feldmeier1998,Neff2003}

Generally speaking, the UCOM and the SRG are two methods to tackle short-range correlations in the nuclear many-body problem by means of unitary transformations. Though both methods start from a different conceptual background---coordinate-space picture of short-range correlations and pre-diagonalization via a flow evolution, respectively---both lead to a decoupling of low-momentum and high-momentum modes. 

\section{Results and discussion}

Here we discuss the interactions and model spaces used for the no-core MCSM and provide some benchmark calculations for the $^{3}$H and $^{4}$He ground states. The model space of the MCSM is spanned by a harmonic oscillator basis truncated with respect to the unperturbed single-particle energies $e_{\rm max} = 2\,n+l$. Our calculations are performed in the model space with $e_{\rm max} = 3$. We use SRG-transformed realistic two-nucleon interactions (hereinafter referred to as $V_{\rm SRG}$) and UCOM-transformed ones ($V_{\rm UCOM}$) as the input potential in the MCSM. And the transformed potentials are derived from the N$^3$LO interaction~\cite{Entem2003,Epelbaum2006,Machleidt}. The Coulomb interaction in all of our calculations is neglected throughout this work for simplicity as discussed in Ref.~\cite{Liu2012}.


\begin{figure}
\includegraphics[width=0.8\textwidth]{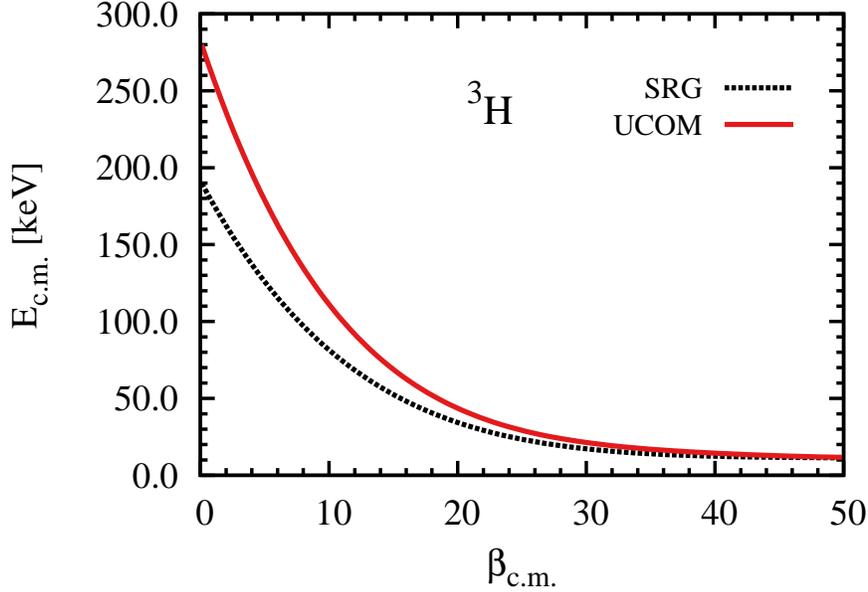}
\caption{\label{fig1} (color online) The center-of-mass motion energies $E_{\rm c.m.}$ of $^{3}$H with $V_{\rm UCOM}$ (solid) and $V_{\rm SRG}$ (dashed) potentials in the MCSM calculations as a function of Lawson's prescription parameter $\beta_{\rm c.m.}$ defined in Eq.~(\ref{eq3_31}). The model space is selected as $e_{\rm max} = 3$. The oscillator parameter $\hbar\,\omega$ is adopted as 28 MeV.}
\end{figure}

As discussed in Sec.~2, the spurious center-of-mass motion should be removed to obtain the intrinsic total energy, if they are mixed in calculated eigenfunctions when different major shells are included in the MCSM calculations. Fig.~\ref{fig1} shows the expectation values $E_{\rm c.m.}$ of $H_{\rm c.m.}$ for $^{3}$H calculated by the MCSM with $V_{\rm SRG}$ (dashed) and $V_{\rm UCOM}$ (solid) in the model space $e_{\rm max} = 3$ as a function of $\beta_{\rm c.m.}$, which is the Lawson's prescription parameter. The harmonic oscillator parameter $\hbar\,\omega$ is adopt as $28$ MeV. The expectation values $E_{\rm c.m.}$ for both $V_{\rm SRG}$ and $V_{\rm UCOM}$ cases decreases rapidly and are less than $50$ keV when $\beta_{\rm c.m.}$ is greater than $20$. In this way, the spurious center-of-mass motion can be suppressed to a large extent by choosing a suitable $\beta_{\rm c.m.}$ value, for instance, $\beta_{\rm c.m.} = 30$. At this point, these two transformed interactions do not make differences with the treatment of spurious center-of-mass motion in the no-core MCSM calculations.


\begin{figure}
\includegraphics[width=0.8\textwidth]{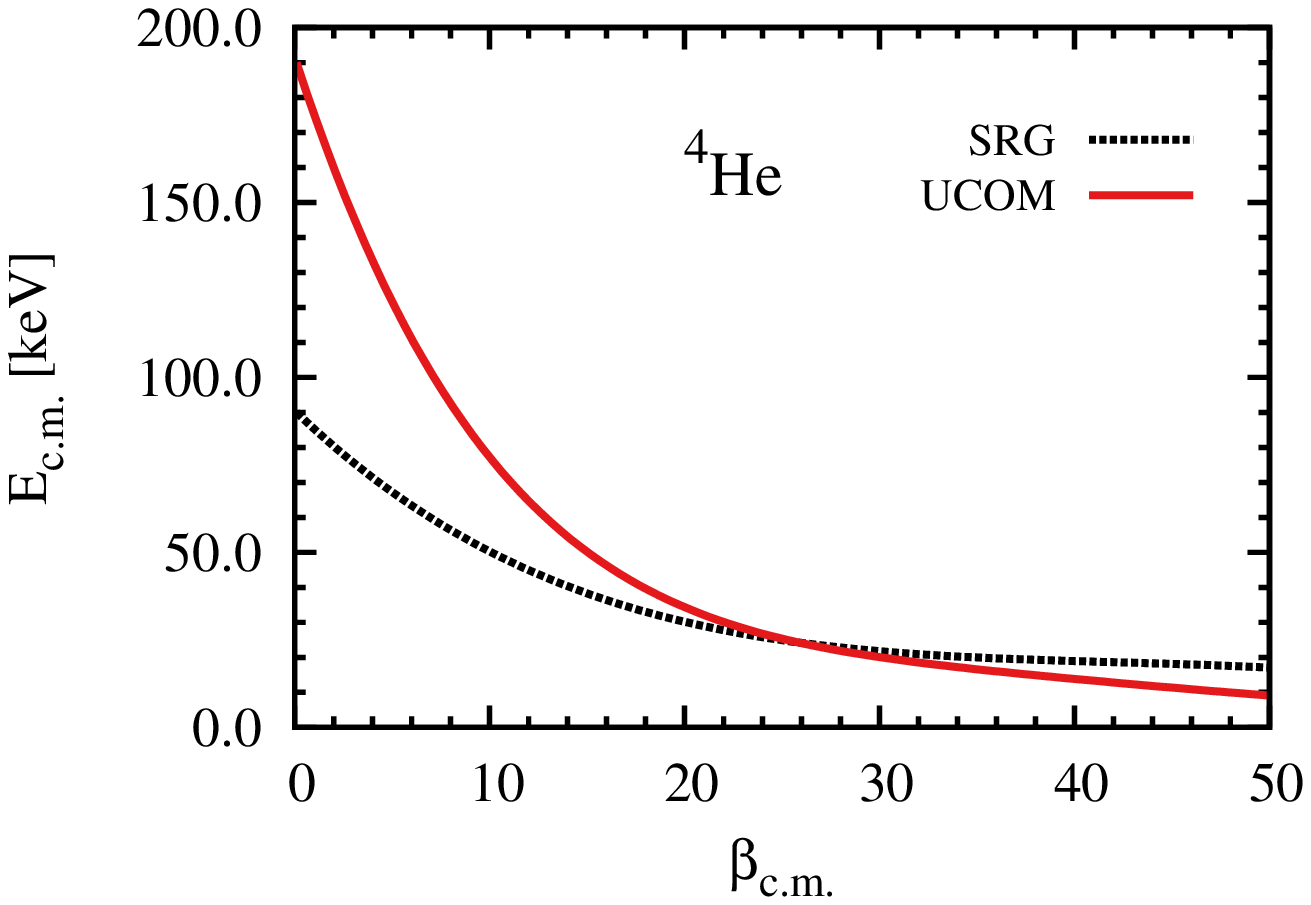}
\caption{\label{fig2} (color online) The center-of-mass motion energies $E_{\rm c.m.}$ of $^{4}$He with UCOM (solid) and SRG (dashed) potentials in the MCSM calculations as a function of Lawson's prescription parameter $\beta_{\rm c.m.}$ defined in Eq.~(\ref{eq3_31}) The model space is selected as $e_{\rm max} = 3$. The oscillator parameter $\hbar\omega$ is adopted as 36 MeV.}
\end{figure}

Figure~\ref{fig2} shows the same expectation values $E_{\rm c.m.}$ of $H_{\rm c.m.}$ but for $^{4}$He calculated by the MCSM with $V_{\rm SRG}$ (dashed) and $V_{\rm UCOM}$ (solid) in the model space $e_{\rm max} = 3$ as a function of $\beta_{\rm c.m.}$. The harmonic oscillator parameter $\hbar\,\omega$ is adopt as $36$ MeV for $^{4}$He. The expectation values $E_{\rm c.m.}$ for both $V_{\rm SRG}$ and $V_{\rm UCOM}$ cases decreases rapidly and are less than $30$ keV when $\beta_{\rm c.m.}$ is greater than $20$. The spurious center-of-mass motion can be treated properly with $\beta_{\rm c.m.} \geqslant 30$ in both cases. Moreover, the $E_{\rm c.m.}$ calculated by MCSM with $V_{\rm UCOM}$ is more close to zero than that with $V_{\rm SRG}$ with $\beta_{\rm c.m.} \geqslant 30$.

\begin{figure}
\includegraphics[width=0.8\textwidth]{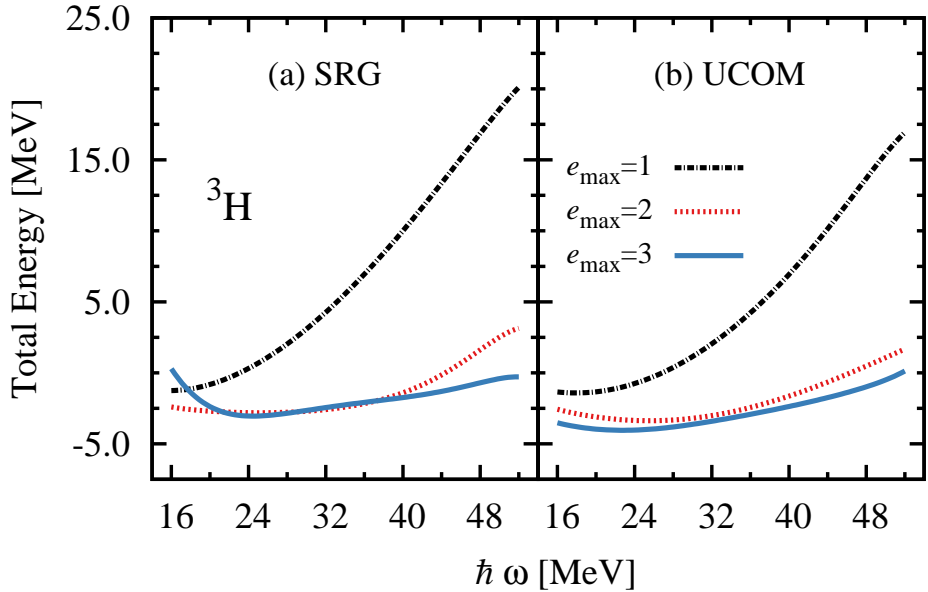}
\caption{\label{fig3} (color online) Calculated total energies of $^{3}$H with (a) SRG and (b) UCOM potentials in the MCSM as a function of oscillator parameter $\hbar\,\omega$ in model spaces $e_{\rm max} =$ 1 (dash dot), 2 (dot) and 3 (solid).}
\end{figure}

With Lawson's prescription parameter $\beta_{\rm c.m.}= 30$, the total intrinsic energy for the ground state of $^{3}$H and $^{4}$He can be evaluated. Likewise, the no-core MCSM calculation results are dependent with the harmonic oscillator parameter $\hbar\,\omega$ due to the truncation of model space like other shell model calculations. In Fig.~\ref{fig3}, the total energies of $^{3}$H calculated by the MCSM with (a) $V_{\rm SRG}$ and (b) $V_{\rm UCOM}$ in model space $e_{\rm max} =$ 1 (dash dot), 2 (dot) and 3 (solid) as a function of harmonic oscillator parameter $\hbar\,\omega$ are shown. The ground state energy for small model spaces, e.g., $e_{\rm max} = 1$, shows a sizable dependence on $\hbar\,\omega$ in both cases. By increasing the size of the model space, the ground state is lowered and dependence on $\hbar\,\omega$ is reduced since the basis in the shell model is close to a complete set. About $4$ MeV of the ground state energy variations are presented for a range of oscillator frequencies $\hbar\,\omega$ from 16 to 52 MeV. However, the MCSM results with $V_{\rm UCOM}$ reveal lower total energy than that with $V_{\rm SRG}$. This differs from the results calculated by the no-core shell model, in which the SRG potential presents lower total energy, or in another words, faster convergence than UCOM potential~\cite{Roth2008}. 

\begin{figure}
\includegraphics[width=0.8\textwidth]{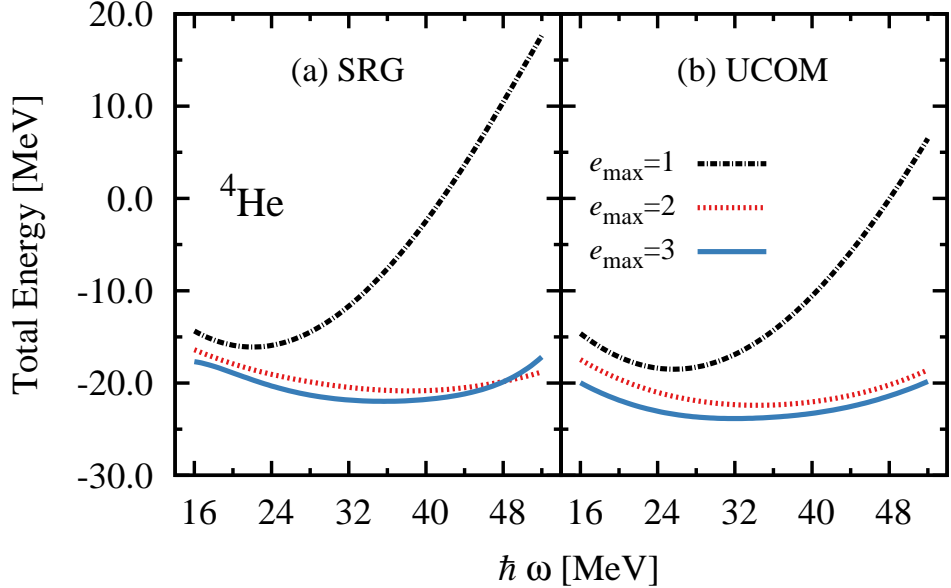}
\caption{\label{fig4} (color online) Calculated total energies of $^{4}$He with (a) SRG and (b) UCOM potentials in the MCSM as a function of oscillator parameter $\hbar\,\omega$ in model spaces $e_{\rm max} =$ 1 (dash dot), 2 (dot) and 3 (solid).}
\end{figure}

In Fig.~\ref{fig4}, the total energies of $^{4}$He calculated by the MCSM with (a) $V_{\rm SRG}$ and (b) $V_{\rm UCOM}$ in model space $e_{\rm max} =$ 1 (dash dot), 2 (dot) and 3 (solid) as a function of harmonic oscillator parameter $\hbar\,\omega$ are also shown. Same properties are presented as shown in Fig.~\ref{fig3} for $^{3}$H. The ground state energy for small model spaces also shows a sizable dependence on $\hbar\,\omega$ in both cases. The ground state is lowered and dependence on $\hbar\,\omega$ is reduced for large model spaces. About $5$ MeV of the ground state energy variations are presented for a range of oscillator frequencies $\hbar\,\omega$ from 16 to 52 MeV. Similarly, the UCOM potentials obtain lower total energy than SRG potentials. The SRG-evolution causes a pre-diagonalization at all momentum scales, i.e. it also leads to a decoupling among the high-$q$ or large-$n$ states. The UCOM-transformed interaction generates a stronger coupling among high-lying states, i.e., the pre-diagonalization in the high-$q$ or large-$n$ regime is not as perfect. The no-core shell model employs the single particle excitation energy truncation scheme. However, the no-core MCSM calculation adopts major shell truncation. The large-$n$ components of SRG-transformed nuclear interaction can be considered to a large extent in no-core shell model, and be omitted somehow in the no-core MCSM. In other words, UCOM-transformed potentials are more suitable than SRG potentials for the MCSM calculations, and SRG potentials show better performance in \textit{ab initio} no-core shell model. 

\section{Summary}

In summary, the UCOM and SRG are compared and discussed in the framework of no-core MCSM for $^{3}$H and $^{4}$He. The treatment of spurious center-of-mass motion by Lawson's prescription are performed in the MCSM calculations. The calculation results with both transformed interactions show good suppression of spurious center-of-mass motion with $\beta_{\rm c.m.}\geqslant 20$. Although both the SRG-evolved and the UCOM-transformed interactions lead to a rapid convergence of NCSM calculations for light nuclei, the UCOM potentials obtain lower total energy than SRG potentials in the MCSM calculations. The SRG-evolution leads to a decoupling among the high-q or large-n states. However, the pre-diagonalization in the high-q or large-n regime is not as perfect as the UCOM-transformed interaction. Hence, the excitation energy truncation in NCSM are proper to the SRG, in which more high-$n$ components can be included. The MCSM with major shell truncation takes more correlations among orbits in one shell and more suitable for UCOM potentials. 

\begin{acknowledgments}
This work is supported by ``the Fundamental Research Funds for the Central Universities"(JUSRP1035), National Natural Science Foundation of China (NSFC) under Grant No. 11305077. The author are indebted to Professor MENG Jie for his suggestions and Professor TAKAHARU Otsuka for inspiration. We are particularly grateful to Professor ROBERT Roth for his support of matrix elements.
\end{acknowledgments}



\end{document}